\documentclass{ws-ijmpe}

\begin{document}

\markboth{B.Nerlo- Pomorska, J.Sykut}{New Set of the Relativistic Mean Field Theory Parameters}

\catchline{}{}{}{}{}
%
%%%%%%%%%%%%%%%%%%%%%%%%%%%%%%%%%%%%%%%%%%%%%%%%%%%%%%%%%%%%%%%%%%%%

\title{A NEW PARAMETER SET FOR THE RELATIVISTIC MEAN FIELD THEORY}

\author{Bo\.zena Nerlo-Pomorska, Joanna Sykut}
\address{Institute of Physics, M. Curie-Sk\l odowska University, 20-031 Lublin, Poland}
\maketitle
\begin{abstract}
Subtracting the Strutinsky shell corrections from the selfconsistent energies
 obtained within
the Relativistic Mean Field Theory (RMFT) we have got estimates for the macroscopic part of the binding energies of 
142 spherical even-even nuclei. By minimizing their root mean 
square deviations from the values obtained with the Lublin-Srasbourg Drop (LSD) model 
with respect to the nine RMFT parameters we have found the optimal set (NL4). The 
new parameters reproduce also the radii of these nuclei with an
accuracy comparable with that obtained with the NL1 and NL3 sets.
\end{abstract}

\section{RMFT parameters}
 The Relativistic  Mean Field Theory (RMFT) \cite{Wal74}, which is essentialy
a  Hartree-Fock like method based on the Dirac equation for nucleons and the
Klein-Gordon equation for the : $\rho$, $\omega$, $\sigma$ mesons and 
photons, reproduces well the nuclear properties
when its parameters (i.e.: masses of nucleons: $m$, mesons: $m_\rho$,
$m_\omega$, $m_\sigma$ and their coupling constants $\rho_\rho$, $\rho_\omega$, 
$\rho_\sigma$,
$\rho_2$, $\rho_3$) are fitted to the largest possible  amount of nuclear data.
Usually the masses and mean square radii of 8 magic nuclei were used
to find the RMFT parameters and  several sets were established for various
regions and quantities as: masses, barriers or radii. We have chequed the quality of  three sets:
NL1 \cite{Rei86}, NL2 \cite{Ruf88} , NL3 \cite{Lal97} by comparing their
 macroscopic
energies  obtained by subtracting the Strutinsky shell correction,
from the  RMFT binding energies evaluated without pairing forces \cite{Ner02}
with the  Lublin-Strasbourg-Drop
(LSD) energy
\cite{Pom03} .
The minimization of the root mean square of the differences of
 macroscopic energy  allowed to
find the new set  of RMFT parameters NL4, which even if rather close to the NL3, 
as can be seen on the table below, results in a rms deviation that is more than 2 times better
(7.17 MeV and 3.29 MeV respectively).
\medskip
\begin{center}
\begin{tabular}{|l|c|c|c|c|c|c|c|c|c|}
\hline
 Set &$m$&$m_\omega$&$m_\rho$&$m_\sigma$&$g_\omega$&$g_\rho$&$g_\sigma$&$\rho_2$&$\rho_3$  \\
\hline
 NL3&939&782.501&763.0&508.194&12.868&--4.474&10.217&--10.431&--28.885 \\
 NL4&938&782.474&763.9&508.194&12.867&--4.360&10.216&--10.432&--28.882 \\
\hline
 \end{tabular} 
\end{center}
\section{Radii}
\begin{figure}[h]
%\vspace{-4.5cm}
\begin{center} \epsfxsize=120mm \epsfbox{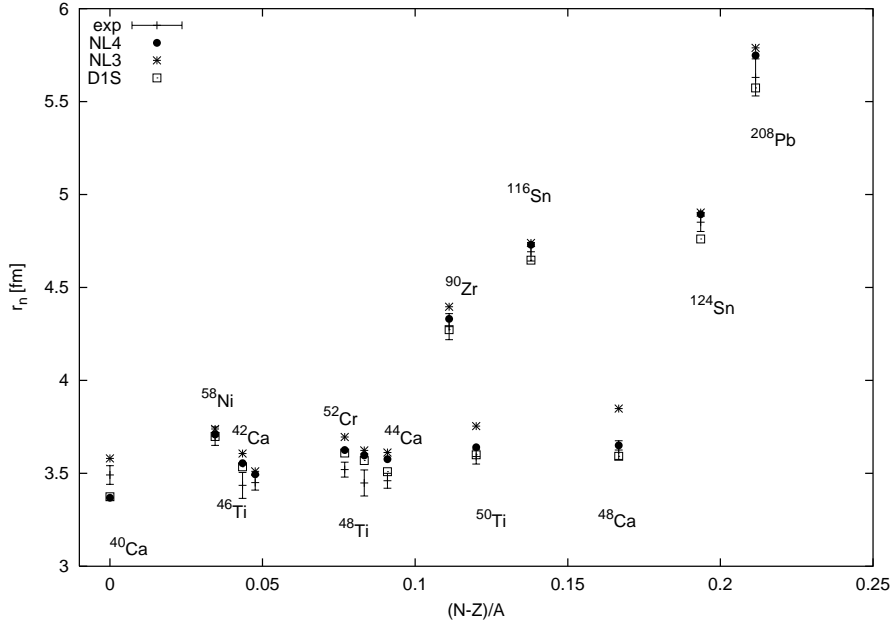} \end{center}
%\vspace{-1.5cm}
\caption{Neutron density radii of 13 spherical , even-even nuclei  evaluated
with the Gogny D1S \protect\cite{Ber84}, and the NL3 \protect\cite{Lal97},  NL4
RMFT parameters sets are compared with the experimental data
\protect\cite{Bat89} .}
\end{figure}

The rms radii of the neutron and charge density distributions obtained with the new  
NL4 set are about as good as those of D1S Gogny \cite{Ber84} force
but systematically better than those of NL3 \cite{Lal97} except for the heaviest isotopes.
 This can be observed in Fig. 1 for 13 neutron and in Fig. 2
for the charge radii. The NL4 results are in best agreement with the experimental
 data.

\begin{figure}[h]
%\vspace{-4.5cm}
\begin{center} \epsfxsize=120mm \epsfbox{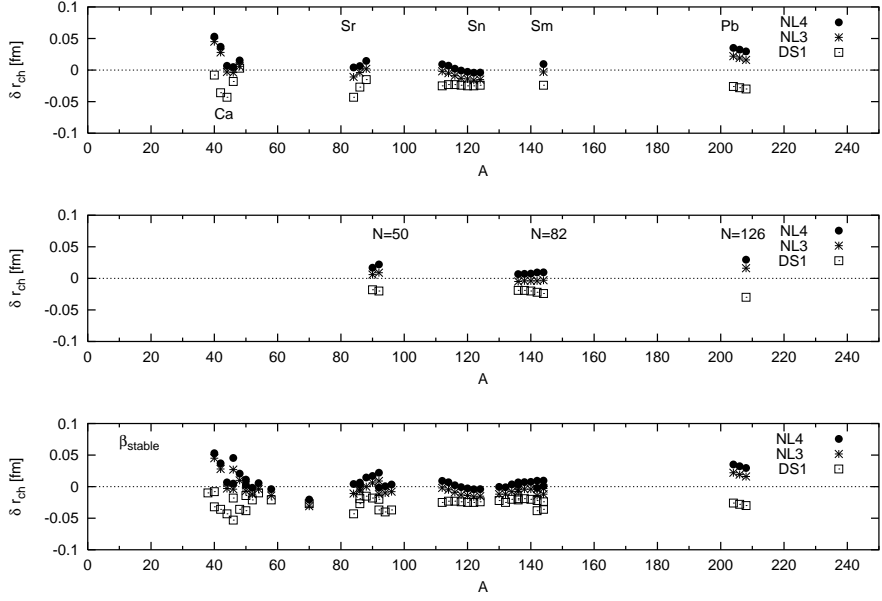} \end{center}
%\vspace{-1.5cm}
\caption{Rms charge radii of spherical even-even nuclei obtained with the Gogny
 D1S force \protect\cite{Ber84} and the NL3 \protect\cite{Lal97}, and NL4
 parameter sets of the RMFT relied  to the experimental data
 \protect\cite{Fri95}, as function of the mass number A for five isotopic
 chains, (upper part) , three isotonic chains  (middle part) and the $\beta$
 stable nuclei with $A>40$ (lower part).}
 \end{figure}

 We found that the charge, neutron and proton radii as well as the ratio 
 {$r_p/r_n$}
obtained within the RMFT with the new NL4 parameter set can be very well
approximated by the expression

\begin{equation}
 r = r_0 \left(1 + \alpha {N - Z\over A} + {\kappa\over A}\right) A^{1/3}
\end{equation}
with respectively parameters listed in the table below. 
\begin{center}
\medskip
\begin{tabular}{|l|c|c|c|}
\hline 
 & $r_0$ &  $\alpha$ &  $\kappa$ \\
\hline
 $r_{ch}$ & 1.2328 fm & --0.15 & 2.1253 \\
 $r_p$   & 1.2257 fm & --0.152& 1.1355 \\
 $r_n$   & 1.1761 fm & 0.2625 & 3.085 \\
 ${r_p\over r_n}$ & 1.0378 & --0.3702 & --1.6249 \\
\hline
\end{tabular}
\end{center}
\begin{figure}[h]
%\vspace{-4.5cm}
\begin{center} \epsfxsize=120mm \epsfbox{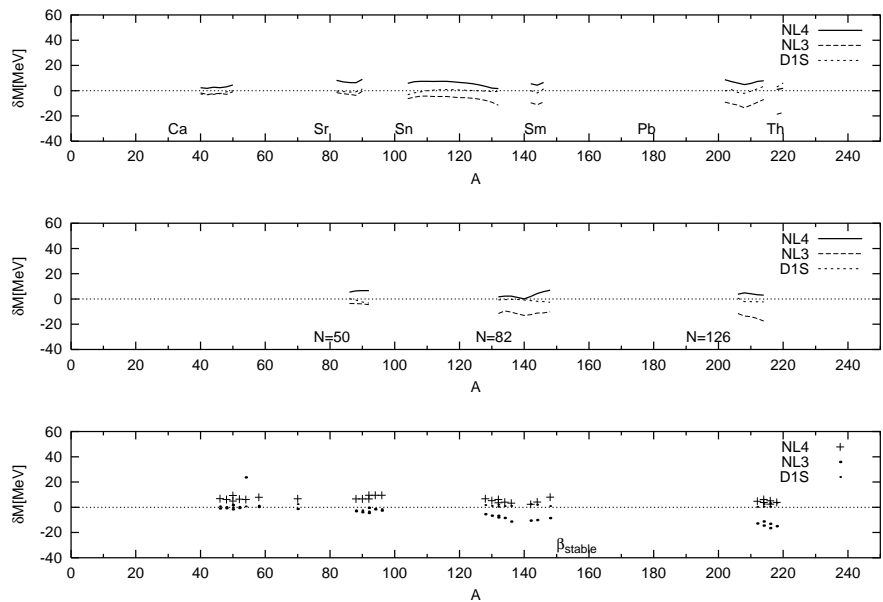} \end{center}
%\vspace{-1.5cm}
\caption{Masses relied to experimental data \protect\cite{Pom03} of 142
spherical , even-even nuclei evaluated with the  Gogny D1S
\protect\cite{Ber84},NL3 \protect\cite{Lal97}, and NL4 RMFT parameters sets  in
function of mass number A,  for isotopes, (upper part) , isotones (middle part)
and $\beta$ stable nuclei (lower part).}
 \end{figure}
\section{Masses}
In Fig. 3 the derivation of the theoretical and experimental masses obtained
 with the RMFT for the NL3 and NL4
 parameter sets of are compared with the Gogny D1S \cite{Ber84}
  Hartree-Fock-Bogolubov (HFB) results.
 As one can see the NL4 masses are closer to the 
experimental data than NL3 ones, though they don't reach the high quality
of the D1S Gogny force.  The 
discrepancies in Fig. 3 are caused by the approximate  treating of the pairing 
force. In RMFT model it is added by a BCS procedure with the experimental
pairing gaps as an input, while the HFB method includes pairing correlations in the selfconsistent
mean field.

%The liquid-drop like formula for the macroscopic part of the RMFT NL4 energy in MeV 
%can be put into the 
%following form :
%\begin{equation}
% E^{\rm macr}_{\rm RMF} = 15.02 (1 - 1.60I^2) A - 16.46 (1 - 1.07I^2) A^{2/3}
%   - 0.68 Z^2/A^{1/3} + 1.13 Z^2/A
%\end{equation}
%These  parameters are close to the LSD formula \cite{Pom03}, the remaining 
%discrepancies origin from the smaller amount of fitted  nuclear data.

\section{Conclusions}

The following conclusions can be drawn from our calculation:
\begin{itemize}
\item[1.] We have shawn that the new RMFT NL4 parameters set reproduces the data of spherical,
even-even nuclei better than the previous ones like the D1S Gogny force.

%\item[2.] The formula for the macroscopic part of the binding energy
%obtained with  the RMFT NL4 set
%is similiar to the old liquid drop ones.
\item[2.] The isospin dependent formulae for radii in RMFT NL4 are closer to
 the adjusted to experimental data ones than for other RMFT forces.
    \end{itemize}

\end{document}